\documentclass{ws-mpla}

\usepackage{graphicx}
\usepackage{dcolumn}
\usepackage{amssymb}
\usepackage{amsmath}
\usepackage{amsfonts}
\usepackage{amsbsy}
\usepackage{color}
\usepackage{rotating}
\usepackage[english]{babel}

\newcommand{\ba}{\begin{eqnarray}}
\newcommand{\ea}{\end{eqnarray}}

\begin{document}

\markboth{V. Salzano, Y. Wang, I. Sendra, R. Lazkoz}
{Linear dark energy equation of state revealed by supernovae?}

\title{Linear dark energy equation of state revealed by supernovae?}

\author{\footnotesize Vincenzo Salzano}
\address{Fisika Teorikoaren eta Zientziaren Historia Saila, Zientzia eta
Teknologia Fakultatea, \\ Euskal Herriko Unibertsitatea, 644 Posta Kutxatila,
48080 Bilbao, Spain\\enzo.salzano@gmail.com}

\author{Yun Wang}
\address{Homer L. Dodge Department of Physics $\&$ Astronomy, Univ. of Oklahoma,\\ 440 W Brooks St., Norman, OK 73019, U.S.A.}

\author{Irene Sendra and Ruth Lazkoz}
\address{Fisika Teorikoaren eta Zientziaren Historia Saila, Zientzia eta
Teknologia Fakultatea, \\ Euskal Herriko Unibertsitatea, 644 Posta Kutxatila,
48080 Bilbao, Spain}

\maketitle

\pub{Received (Day Month Year)}{Revised (Day Month Year)}

\begin{abstract}
In this work we propose a test to detect the linearity of the dark energy equation of state, and apply it to the SNLS3 Type Ia Supernova (SN Ia) data set. We find that: \textit{a.}
current SN Ia data are well described by a dark energy equation of state linear in the cosmic scale factor $a$, at least up to a redshift $z =1$, independent of the pivot points chosen for the linear relation;
\textit{b.} there is no significant evidence of any deviation from linearity. This apparent linearity may reflect the limit of dark energy information extractable from current SN Ia data.

\end{abstract}

\ccode{PACS Nos.: include PACS Nos.}

\section{Introduction}

After the discovery of the accelerated expansion of our Universe \cite{SNfirst1,SNfirst2}, many models have been proposed to solve the mystery of dark energy, both in the context of general relativity and
of alternative gravity theories \cite{alttheo1,alttheo2,alttheo3,alttheo4,alttheo5,alttheo6,alttheo7,alttheo8,alttheo9,alttheo10,alttheo11,alttheo12,alttheo13,alttheo14,alttheo15,alttheo16,alttheo17}. Models are characterized by an analytical expression for the dark energy equation of state (EoS), derived either by phenomenological or theoretical considerations;
and generally depending on two or more parameters \cite{twopar1,twopar2} which can be constrained using observations. A very short, not exhaustive, summary of the most used EoS parametrizations is in \cite{CPL1,CPL2,eosmodel1,eosmodel2,eosmodel3,eosmodel4,eosmodel5,eosmodel6,eosmodel7,eosmodel8,eosmodel9,eosmodel10};
 among them, the CPL model \cite{CPL1,CPL2} is considered as the reference model. It is defined as
$w(a) = w_{0} + (1-a) w_{a}$: $w_{0}$ is the EoS present value $(a=1$ or $z = 0)$ and $w_{\infty} = w_{0} + w_{a}$ is the asymptotic value of EoS at early times $(a \rightarrow 0$ or $z\rightarrow \infty)$.
This model has some well known problems: the high correlation between its two parameters, and the high-redshift dependence of the parameter $w_{a}$, which makes its use with low-redshift limited data questionable.
An alternative parametrization is \textit{Wang's model} \cite{Wang2008}:
\begin{equation}\label{eq:wang}
w(a)=  \left( \frac{a_{c} - a}{a_{c} - 1} \right) w_{0} + \left( \frac{a - 1}{a_{c} - 1} \right) w_{c} \; ,
\end{equation}
where $w_{c} = w(a_{c})$ is the EoS calculated at the pivot value for the cosmic scale factor $a$, chosen to minimize the correlation between the EoS parameters, $w_{0}$ and $w_{c}$. In a broader view, we can see that both CPL and Wang's model are linear interpolations between two points, $(a_{i}, w_{i} = w(a_{i}))$ and $(a_{j}, w_{j} = w(a_{j}))$; thus we can define a \textit{General Linear EoS model} (GL)
\begin{equation}\label{eq:linear}
w(a) = \left( \frac{a_{j} - a}{a_{j} - a_{i}} \right) w_{i} + \left( \frac{a_{i} - a}{a_{i} - a_{j}} \right) w_{j} \; .
\end{equation}
We obtain the CPL model if $a_{i} = a_{0} = 1$ and $a_{j} = 0$, i.e. $w_j=w_{\infty}$; just note that in the CPL model $w_{\infty}=w_0+w_a$. Similarly, Eq.~(\ref{eq:wang}) is recovered if $a_{i} = a_{0} = 1$ and $a_{j} = a_{c}$.

GL models can be used to test whether the EoS is \textit{linear or not} in the scale factor. This test has nontrivial implications: if linearity is verified, all models in \cite{eosmodel} deviating from it may be discarded. If the EoS is not effectively linear, then the CPL model is \textit{intrinsically wrong}. To test this possibility in the most general way, we can generalize the GL model to higher-order polynomial functions; an interpolation function between $\mathcal{N}$ points can be written in a general fashion as:
\begin{equation}\label{eq:genlinear}
w(a) = \sum^{\mathcal{N}}_{i=1} w_{i} \left(\prod^{\mathcal{N}}_{j\neq i} \frac{a_{j} - a}{a_{j} - a_{i}} \right)
\end{equation}
with $\mathcal{N}=2$ for the GL model; $\mathcal{N}= 3$ for a \textit{second-order interpolation model} $(2NL)$; $\mathcal{N}= 4$ for a \textit{third-order interpolation model} $(3NL)$; and so on. A third possibility is that current SN Ia data do not allow us to distinguish between a linear and a nonlinear EoS, and linearity could be an \textit{artifact} coming from the data used.

\begin{figure*}
\centering
\includegraphics[width=7.5cm]{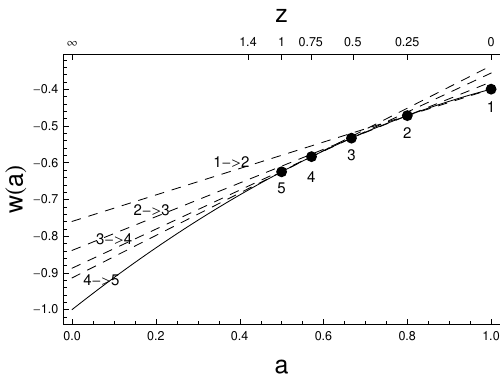}
\\
~~~
\\
\includegraphics[width=7.5cm]{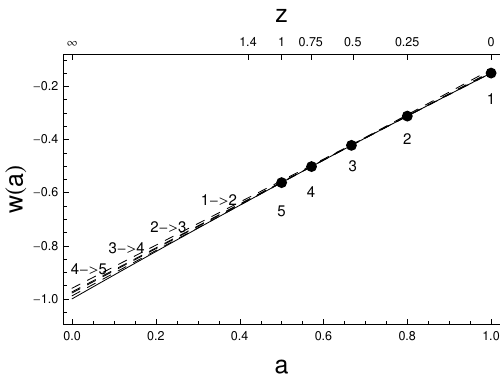}
\caption{Test of linearity of dark energy EoS through the GL models defined in Eq.~(\ref{eq:linear}): solid line is the underlying unknown EoS function which we want to reconstruct; dots are the pivot scale factor values chosen to define various GL models; dashed lines are the GL models calculated varying the pivot values. In the left panel, the underlying EoS is not linear; in the right one, it is almost linear (very small curvature).}
\label{fig:linear_test}
\end{figure*}

This test can be implemented very easily, as illustrated in Fig.~\ref{fig:linear_test}\footnote{Fig.~\ref{fig:linear_test} is purely illustrative: we are ``not'' assuming that the error on $w(a)$ is minimized at the pivot points. The choices for Fig.~\ref{fig:linear_test} are made only for illustrating in the clearest way the method we are going to employ; updating it with errors at the pivot points would leave the scope and the results unchanged. The errors and discussion about their weight in the analysis will be taken into account in the quantitative analysis we will show in next sections}. Let us assume an exact EoS with the (unknown) profile given by the solid line in the figure; the dots are the pivot values where the GL EoS is to be calculated; the dashed lines represent the GL models (linear interpolations) using different pairs of pivots. If the underlying EoS is not linear (left panel), then we will obtain different estimations of the pivot values when using different GL parameterizations; on the other hand, if the EoS is linear (or with a \textit{very small} curvature; right panel), such estimations will yield the same linear relation, independent of the GL model assumed.

We want to stress here that our main scope is to confirm or confute definitely if a linear fit of $w(a)$ for the dark energy equation of state is applicable to current data or not. \textit{IF AND ONLY IF} we find a particular trend in the estimations of the $w(a)$ pivot values when we change the parametrization, \textit{THEN} we can conclude that the dark energy equation of state can be parameterized by a linear model, and only \textit{AFTER} this result we can assert that changing the pivot values only corresponds to a re-parametrization of the same (linear) model. \textit{OTHERWISE}, the dark energy equation of state is not linear and each one of the models defined as Generalized linear (GL), by changing the pivot values, are independent parameterizations (see left panel of our Fig.~\ref{fig:linear_test}).

\section{Data}

We will use SN Ia data from the SNLS3 compilation that includes the three year data from the \textit{SuperNova Legacy Survey} \cite{SNLS1,SNLS2,SNLS3}. The $\chi^2$ is generally defined as $\chi^2 = \Delta \boldsymbol{\mathcal{F}} \; \cdot \; \mathbf{C}^{-1} \; \cdot \; \Delta  \boldsymbol{\mathcal{F}}$. $\Delta\boldsymbol{\mathcal{F}} = \mathcal{F}_{theo} - \mathcal{F}_{obs}$ is the difference between the observed and theoretical value of the observable quantity, $ \mathcal{F}$; this will be the SN Ia magnitude $m_{mod}$ for SNLS3:
\begin{equation}
\label{eq:m_snls3}
m_{\rm mod} = 5 \log_{10} [H_0 d_{L}(z, \Omega_m; \boldsymbol{\theta}) ] - \alpha (s-1) + \beta \mathcal{C} + \mathcal{M} \; .
\end{equation}
Note that
\ba
\label{eq:dl_H}
&&H_0 d_{L}(z, \Omega_m; \boldsymbol{\theta}) = (1+z) \ \int_{0}^{z}
\frac{\mathrm{d}z'}{E(z', \Omega_m; \boldsymbol{\theta})} \, ,\\
&&E(z) = \frac{H(z)}{H_0}=\left[ \Omega_{m} (1+z)^{3}+ (1-\Omega_{m})X(z,\boldsymbol{\theta}) \right]^{1/2} \, ,\\
&&X(z,\boldsymbol{\theta})= \exp \left[3\int_{a}^{1} \frac{da'}{a'} (1+w(a',\boldsymbol{\theta})) \right] \, .
\ea
with $\boldsymbol{\theta}$ the dark energy EoS parameters vector ($\boldsymbol{\theta} = (w_{i}, w_{j})$ for GL models) and spatial flatness being assumed.

We apply a Gaussian prior on the matter content, $\Omega_{m} = 0.26 \pm 0.02$ \cite{Wang11}, to reduce the degeneracy among the EoS dark energy parameters and include information from external cosmological datasets other than SN Ia assuming a two parameter $w(z)$ model and arbitrary curvature. $\mathcal{M}$ is a nuisance parameter combining the Hubble constant $H_{0}$ and the absolute magnitude of a fiducial SN Ia, and we minimized $\chi^2$ (through a Monte Carlo Markov Chain (MCMC) algorithm) by marginalizing over it \cite{SNLS}. Finally, $\mathbf{C}$ is the covariance matrix, which depends on the parameters $\alpha$ and $\beta$, considered as free fitting parameters.
To test each model, we calculate the Bayesian evidence, defined as the probability of the data $D$ given the model $M$ with a set of parameters $\boldsymbol{\theta}$, $\mathcal{E}(M) = \int \mathrm{d}\boldsymbol{\theta} L(D|\boldsymbol{\theta},M)\pi(\boldsymbol{\theta}|M)$:
$\pi(\boldsymbol{\theta}|M) $ is the prior on the set of parameters, normalized to unity, and $L(D|\boldsymbol{\theta},M)$ is the likelihood function. We have been very careful in imposing priors: we impose flat priors on the estimated dark energy parameters over sufficiently wide ranges so that further increasing these ranges has no impact on the results; $\Omega_{m}$ is sampled from a gaussian probability distribution so that, by definition, it is automatically normalized to unity. The evidence is estimated using the algorithm in \cite{evidence}. To reduce the statistical noise we run the algorithm many times obtaining a distribution of $\sim 1000$ values from which we extract the best value of the evidence as the mean of such distribution. Then, we calculate the Bayes Factor, defined as the ratio of evidences of two models, $M_{i}$ and $M_{j}$, $\mathcal{B}_{ij} = \mathcal{E}_{i}/\mathcal{E}_{j}$. If $\mathcal{B}_{ij} > 1$, model $M_i$ is preferred over $M_j$, given the data. We will use the $(0-0.5)$ GL model as reference model $j$. The Bayesian evidence may be interpreted using Jeffreys' Scale \cite{jeffreys}; but in a recent paper \cite{nesseris}, it is shown that the Jeffreys' scale is not a reliable tool for model comparison but at the same time does not question the statistical validity of the Bayes factor as a efficient model-comparison tool: a Bayes factor $\mathcal{B}_{ij}>1$ unequivocally states that the model $i$ is more likely than model $j$. This is why in our analysis we only rely on the values of the Bayes factors.

\section{Results}

There are many criteria that should be followed in order to decide if a model is the preferred one; here they are the criteria we have considered most important, ranked in decreasing importance order:
\begin{itemize}
  \item the Bayes factor
  \item the consistency/variation of each parameter with respect the others
  \item comparison of this variation to the related errors
\end{itemize}
We are going to discuss all these criteria in detail and their consequences; we will also argue the possible sources of each one and the way they can be improved in the future. It is the combination of all these that will allow us to assess the \textit{statistical validity of one model (linear) with respect to another (non-linear)}.

First we focus on the information that can be extracted using different GL models; results are summarized in Table~\ref{tab:bestSNLS1}. We have chosen $5$ pivot values for the scale factor, equally spaced in the redshift space (an arbitrary choice which has no effect on final results) and up to a maximum redshift $z = 1$ (the redshift at which the number of observed SNe Ia is statistically significant). They are $z = \{0, 0.25, 0.5, 0.75, 1\}$ (corresponding to scale factors $a = \{1, 0.8, 0.67, 0.57, 0.5$\}). For each GL parametrization we have two primary parameters (bold text in tables), the EoS calculated at the pivot redshifts and directly derived as fitting parameters; and secondary parameters (plain text), the EoS calculated in the remaining chosen pivot values and derived from the obtained bestfit GL EoS relation. We also perform fits with a cosmological constant, a quiessence and the CPL model for comparison.

\begin{figure*}
\centering
\includegraphics[width=7.5cm]{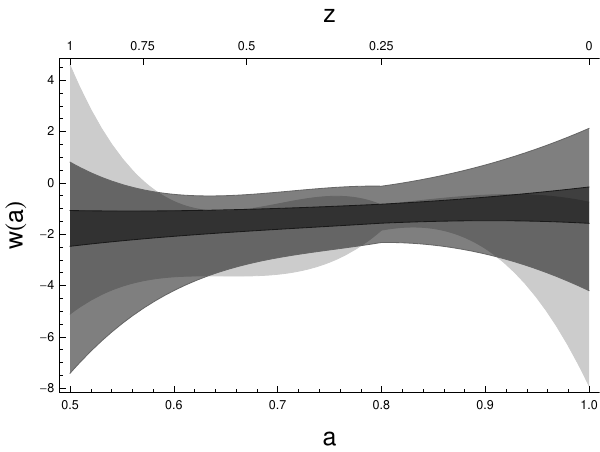}
\caption{Total $68\%$ confidence levels from GL, $2NL$ and $3NL$, respectively from darker to lighter regions.}
\label{fig:summary}
\end{figure*}

We find the following about the absolute value of the EoS parameters:
\begin{itemize}
 \item the primary EoS parameters are almost perfectly consistent with each other, independent of the GL model assumed. We note that, given the nature of the MCMC algorithm, there is an intrinsic statistical background noise (i.e. nonphysically meaningful fluctuations) to be considered;
 \item the same good agreement is valid for secondary parameters; particularly impressive is the consistency with the $w_{\infty} = w_{0} + w_{a}$ derived from the CPL model.
\end{itemize}
These are two strong points in favor of a linear (in scale factor) parametrization for the dark energy EoS, \textit{at least up to $z=1$}. Of course, we have to compare such results with the errors on these parameters:
\begin{itemize}
 \item we join all GL models: for each GL parametrization, we randomly extract $N=1000$ pairs of pivot EoS parameters, $(w_{i}, w_{j})$, from Gaussian distributions centered on the best fit values from each model and with dispersion equal to the MCMC derived errors; we show the total errors on each EoS parameter in the last row of Table~\ref{tab:bestSNLS1} and in darkest region in Fig.~\ref{fig:summary}. We see there is a minimum in dark energy EoS parameters error at $z \sim 0.25$; this same value is obtained in \cite{Wang2008} as the redshift corresponding to the minimal correlation between the pivot parameters, while errors on $w_{0}$ are comparable with errors on $w_{1}$;
 \item the errors generally grow when moving to higher redshift pivot values: this is as expected since the number of observed SNe Ia drops toward higher redshifts. If we consider $0.1$-width redshift bins, the SNLS3 sample has $\sim 120$ objects at $z<0.1$ and $\approx 40$ objects per bin up to $z\sim 0.9$; this, of course, influences the estimation of the pivot EoS parameters with a degradation of the reconstructed value at higher redshifts;
\end{itemize}
Finally, we note that the Bayes factor is $\approx 1$: GL models are practically equivalent. This is an important consistency check in favor of linearity. But, considering the behavior of the errors, an important question arises: \textit{does the error-redshift relation depend on the number of data points, or is it an indication of the intrinsic breakdown of the linear EoS model?}. As we have shown in the left panel of Fig.~\ref{fig:linear_test}, a non-linear EoS can produce different estimated values for the pivot parameters \textit{or}, \textit{equivalently}, a wider dispersion, i.e. larger errors on them. In order to check this, we have analyzed the $2NL$ and $3NL$ models, to be compared with linear models. Results are given in Table.~\ref{tab:bestSNLS2}:
\begin{itemize}
  \item the primary EoS parameters are not as consistent with each other as in the case of GL models; in particular the agreement among the derived $w_{\infty}$ values goes down;
  \item errors are much larger than the GL models, due to the larger number of estimated parameters (see lighter regions in Fig.~\ref{fig:summary}.);
  \item interestingly, even with these models, the $w_{0.25}$ pivot corresponds to the best estimation of the dark energy EoS; this could be a further indication that this depends on the higher number of points at lower redshift more than on intrinsic properties of EoS;
  \item we also note the value of the Bayes Factor: it is $\approx 0.6$ for the $2NL$ models and $\approx 0.58$ for the $3NL$ ones, smaller than the values from the GL models by $\sim 40 \%$, so that non-linear models are clearly disfavored.
\end{itemize}
All the points above lead to the conclusion that a linear EoS for dark energy is the most statistically probable choice when using SNe Ia to probe cosmology. Clearly the values of the errors, as explained above, still leave open the possibility for non-linear models.
With the addition of more SNe Ia in the future and at higher redshifts, we will be able to obtain even stronger constraints on the global EoS trend, and the evidence for linearity of dark energy EoS might strengthen.

To conclude, we have found \textit{that current SN Ia data are well described by a dark energy EoS linear in $a$, independent of
the pivot points chosen for the linear relation, and that there is no strong and significant evidence of any deviation from linearity}.
This may indicate that the dark energy EoS is a linear function in $a$, or that current data only allow the extraction
of a linear function. Significantly larger SN Ia data sets will be required to clarify this.

\section*{Acknowledgments}

V.S., I.S. and R.L. are supported by the Spanish Ministry of Economy and Competitiveness through research projects FIS2010-15492 and Consolider EPI CSD2010-00064, by the Basque Government through the special research action KATEA and ETORKOSMO, and by the University of the Basque Country UPV/EHU under program UFI 11/55. Y.W. is supported in part by DOE grant DE-FG02-04ER41305.

{\renewcommand{\tabcolsep}{2.mm}
{\renewcommand{\arraystretch}{2.5}
\begin{table*}
\begin{minipage}{\textwidth}
\caption{Linear EoS test with SNLS3 data. For any EoS parametrization (first column) we give primary fitted parameters in bold text and secondary derived parameters in plain text. Errors for the primary parameters are derived directly from the MCMC procedure; the errors on secondary parameters are calculated following the standard error propagation theory.}\label{tab:bestSNLS1}
\resizebox*{\textwidth}{!}{
\begin{tabular}{c|ccccccc|cc|c}
\hline
\hline
\textbf{SNLS3} & $\Omega_{m}$ & \multicolumn{6}{c|}{$w$} & $\alpha$ & $\beta$ & $\mathcal{B}_{ij}$\\
\hline
\hline
$\Lambda$CDM   & $0.228^{+0.039}_{-0.038}$ & \multicolumn{6}{c|}{-1 \; (fixed)} & $1.434^{+0.110}_{-0.105}$ & $3.265^{+0.111}_{-0.109}$ & $1.311$ \\
\hline
Quiessence   & $0.257^{+0.020}_{-0.019}$ & \multicolumn{6}{c|}{$-1.049^{+0.106}_{-0.118}$} & $1.429^{+0.113}_{-0.102}$ & $3.260^{+0.112}_{-0.107}$ & $0.740$ \\
\hline
\hline
               & $\Omega_m$ & $w_{0}$ & $w_{0.25}$ & $w_{0.5}$ & $w_{0.75}$ & $w_{1}$ & $w_{\infty}$ & $\alpha$ & $\beta$ & $\mathcal{B}_{ij}$\\
\hline
\hline
CPL              & $0.265^{+0.020}_{-0.020}$ & $\mathbf{-0.827^{+0.215}_{-0.214}}$ & $-1.212^{+0.399}_{-0.422}$ & $-1.469^{+0.617}_{-0.659}$ & $-1.652^{+0.782}_{-0.835}$ & $-1.789^{+0.907}_{-0.970}$ & $\mathbf{-1.925^{+1.587}_{-1.714}}$ & $1.431^{+0.111}_{-0.105}$ & $3.271^{+0.112}_{-0.109}$ & $0.976$\\
\hline
$(0-0.25)$       & $0.265^{+0.020}_{-0.020}$ & $\mathbf{-0.816^{+0.223}_{-0.213}}$ & $\mathbf{-1.215^{+0.180}_{-0.220}}$ & $-1.481^{+0.335}_{-0.393}$ & $-1.671^{+0.462}_{-0.531}$ & $-1.813^{+0.561}_{-0.636}$ & $\mathrm{-2.811^{+1.267}_{-1.391}}$  & $1.434^{+0.113}_{-0.107}$ & $3.276^{+0.111}_{-0.109}$ & $0.952$\\
\hline
$(0-0.5)$       & $0.265^{+0.019}_{-0.020}$ & $\mathbf{-0.824^{+0.220}_{-0.214}}$ & $-1.200^{+0.229}_{-0.279}$ & $\mathbf{-1.451^{+0.352}_{-0.442}}$ & $-1.630^{+0.457}_{-0.572}$ & $-1.764^{+0.539}_{-0.672}$ & $-2.705^{+1.144}_{-1.393}$  & $1.433^{+0.108}_{-0.104}$ & $3.269^{+0.113}_{-0.108}$ & $1.000$\\
\hline
$(0-0.75)$       & $0.265^{+0.020}_{-0.020}$ & $\mathbf{-0.827^{+0.221}_{-0.207}}$ & $-1.203^{+0.254}_{-0.312}$ & $-1.454^{+0.378}_{-0.489}$ & $\mathbf{-1.633^{+0.482}_{-0.626}}$ & $-1.767^{+0.564}_{-0.731}$ & $-2.708^{+1.163}_{-1.486}$ & $1.431^{+0.114}_{-0.102}$ & $3.273^{+0.112}_{-0.110}$ & $0.966$\\
\hline
$(0-1.0)$       & $0.265^{+0.020}_{-0.020}$ & $\mathbf{-0.818^{+0.217}_{-0.215}}$ & $-1.213^{+0.272}_{-0.321}$ & $-1.477^{+0.404}_{-0.494}$ & $-1.665^{+0.512}_{-0.630}$ & $\mathbf{-1.806^{+0.596}_{-0.734}}$ & $-2.794^{+1.212}_{-1.484}$ & $1.432^{+0.111}_{-0.106}$ & $3.275^{+0.11}_{-0.112}$ & $0.961$\\
\hline
$(0.25-0.5)$       & $0.265^{+0.020}_{-0.020}$ & $-0.844^{+0.678}_{-0.866}$ & $\mathbf{-1.210^{+0.173}_{-0.220}}$ & $\mathbf{-1.454^{+0.348}_{-0.446}}$ & $-1.628^{+0.609}_{-0.780}$ & $-1.759^{+0.812}_{-1.040}$ & $-2.674^{+2.260}_{-2.893}$  & $1.436^{+0.111}_{-0.104}$ & $3.271^{+0.112}_{-0.110}$ & $0.972$ \\
\hline
$(0.25-0.75)$      & $0.266^{+0.020}_{-0.021}$ & $-0.837^{+0.547}_{-0.708}$ & $\mathbf{-1.215^{+0.178}_{-0.226}}$ & $-1.467^{+0.299}_{-0.389}$ & $\mathbf{-1.647^{+0.496}_{-0.648}}$ & $-1.782^{+0.653}_{-0.853}$ & $-2.727^{+1.792}_{-2.337}$  & $1.429^{+0.113}_{-0.103}$ & $3.275^{+0.112}_{-0.108}$ & $0.928$ \\
\hline
$(0.25-1.0)$       & $0.265^{+0.020}_{-0.020}$ & $-0.824^{+0.565}_{-0.536}$ & $\mathbf{-1.207^{+0.172}_{-0.219}}$ & $-1.462^{+0.339}_{-0.289}$ & $-1.644^{+0.558}_{-0.452}$ & $\mathbf{-1.781^{+0.589}_{-0.731}}$ & $-2.738^{+1.970}_{-1.612}$  & $1.434^{+0.110}_{-0.102}$ & $3.273^{+0.110}_{-0.109}$ & $1.001$ \\
\hline
$(0.5-0.75)$       & $0.265^{+0.019}_{-0.020}$ & $-0.827^{+2.280}_{-2.924}$ & $-1.201^{+1.060}_{-1.365}$ & $\mathbf{-1.450^{+0.340}_{-0.443}}$ & $\mathbf{-1.628^{+0.483}_{-0.611}}$ & $-1.761^{+0.883}_{-1.120}$ & $-2.696^{+3.949}_{-5.036}$  & $1.438^{+0.113}_{-0.104}$ & $3.273^{+0.109}_{-0.110}$ & $0.986$ \\
\hline
$(0.5-1.0)$       & $0.265^{+0.020}_{-0.020}$ & $-0.839^{+1.539}_{-2.136}$ & $-1.205^{+0.768}_{-1.065}$ & $\mathbf{-1.449^{+0.343}_{-0.474}}$ & $-1.623^{+0.358}_{-0.499}$ & $\mathbf{-1.754^{+0.572}_{-0.797}}$ & $-2.669^{+2.509}_{-3.491}$  & $1.436^{+0.112}_{-0.104}$ & $3.276^{+0.112}_{-0.110}$ & $0.954$ \\
\hline
$(0.75-1.0)$      & $0.265^{+0.020}_{-0.020}$ & $-0.801^{+4.987}_{-6.216}$ & $-1.207^{+2.824}_{-3.521}$ & $-1.478^{+1.401}_{-1.749}$ & $\mathbf{-1.671^{+0.493}_{-0.617}}$ & $\mathbf{-1.816^{+0.600}_{-0.745}}$ & $-2.831^{+5.912}_{-7.360}$  & $1.431^{+0.107}_{-0.104}$ & $3.271^{+0.112}_{-0.107}$ & $0.977$ \\
\hline
\hline
all & $0.265^{+0.073}_{-0.073}$ & $-0.829^{+0.662}_{-0.754}$ & $-1.209^{+0.372}_{-0.368}$ & $-1.455^{+0.417}_{-0.436}$ & $-1.649^{+0.547}_{-0.540}$ & $-1.763^{+0.679}_{-0.714}$ & $-2.742^{+2.103}_{-2.165}$  & $-$ & $-$ & $0.967$
 \\
\hline
\hline
\end{tabular}}
\end{minipage}
\end{table*}}}

{\renewcommand{\tabcolsep}{2.mm}
{\renewcommand{\arraystretch}{2.5}
\begin{table*}
\begin{minipage}{\textwidth}
\caption{Non-linear EoS test with SNLS3 data: column description is the same of Table.~\ref{tab:bestSNLS1}. Asterisks are for non-converged MCMC.}\label{tab:bestSNLS2}
\resizebox*{\textwidth}{!}{
\begin{tabular}{c|ccccccc|cc|c}
\hline
\hline
\textbf{SNLS3} & $\Omega_{m}$ & $w_{0}$ & $w_{0.25}$ & $w_{0.5}$ & $w_{0.75}$ & $w_{1}$ & $w_{\infty}$ & $\alpha$ & $\beta$ & $\mathcal{B}_{ij}$\\
\hline
\hline
$(0-0.25-0.5)$     & $0.270^{+0.021}_{-0.020}$ & $\mathbf{-1.038^{+0.518}_{-0.636}}$ & $\mathbf{-1.224^{+0.258}_{-0.278}}$ & $\mathbf{-1.729^{+0.557}_{-1.383}}$ & $-2.276^{+1.700}_{-2.635}$ & $-2.789^{+2.673}_{-4.091}$ &
$-8.826^{+15.467}_{-15.958}$  & $1.439^{+0.110}_{-0.105}$ & $3.285^{+0.117}_{-0.110}$ & $0.601$ \\
\hline
$(0-0.25-0.75)$     & $0.270^{+0.021}_{-0.021}$ & $\mathbf{-1.007^{+0.515}_{-0.655}}$ & $\mathbf{-1.234^{+0.254}_{-0.275}}$ & $-1.776^{+0.573}_{-0.863}$ & $\mathbf{-2.354^{+1.415}_{-3.399}}$ & $-2.892^{+1.837}_{-2.836}$ &
$-9.170^{+10.583}_{-10.881}$  & $1.436^{+0.110}_{-0.106}$ & $3.290^{+0.114}_{-0.108}$ & $0.620$ \\
\hline
$(0-0.25-1.0)$      & $0.270^{+0.021}_{-0.021}$ & $\mathbf{-1.022^{+0.520}_{-0.573}}$ & $\mathbf{-1.222^{+0.260}_{-0.283}}$ & $-1.815^{+0.539}_{-0.646}$ & $-2.465^{+1.579}_{-1.944}$ & $\mathbf{-3.075^{+2.493}_{-3.781}}$ & $-10.305^{+8.940}_{-9.040}$  & $1.433^{+0.112}_{-0.102}$ & $3.284^{+0.115}_{-0.111}$ & $0.610$ \\
\hline
$(0-0.5-0.75)$     & $0.272^{+0.021}_{-0.022}$ & $\mathbf{-1.078^{+0.564}_{-0.656}}$ & $-1.225^{+1.279}_{-2.507}$ & $\mathbf{-1.863^{+0.672}_{-1.519}}$ & $\mathbf{-2.583^{+1.606}_{-3.474}}$ & $-3.268^{+2.696}_{-4.172}$ &
$-11.530^{+24.021}_{-34.338}$  & $1.438^{+0.110}_{-0.106}$ & $3.290^{+0.116}_{-0.112}$ & $0.546$ \\
\hline
$*(0-0.5-1.0)$      & $0.276^{+0.021}_{-0.023}$ & $\mathbf{-1.281^{+0.682}_{-1.029}}$ & $-1.183^{+1.193}_{-3.890}$ & $\mathbf{-2.208^{+0.933}_{-3.450}}$ & $-3.474^{+1.077}_{-2.531}$ & $\mathbf{-4.716^{+3.728}_{-11.625}}$ & $-20.418^{+17.634}_{-34.741}$  & $1.438^{+0.114}_{-0.104}$ & $3.303^{+0.115}_{-0.112}$ & $0.458$ \\
\hline
$*(0-0.75-1.0)$    & $0.277^{+0.022}_{-0.022}$ & $\mathbf{-1.338^{+0.676}_{-1.026}}$ & $-1.244^{+5.225}_{-16.415}$ & $-2.376^{+4.593}_{-14.942}$ & $\mathbf{-3.768^{+2.349}_{-8.050}}$ & $\mathbf{-5.132^{+3.698}_{-12.448}}$ & $-22.352^{+49.189}_{-135.043}$  & $1.441^{+0.110}_{-0.108}$ & $3.302^{+0.117}_{-0.108}$ & $0.440$ \\
\hline
$(0.25-0.5-0.75)$      & $0.269^{+0.021}_{-0.020}$ & $-1.118^{+5.270}_{-11.106}$ & $\mathbf{-1.229^{+0.263}_{-0.290}}$ & $\mathbf{-1.745^{+0.544}_{-1.416}}$ & $\mathbf{-2.330^{+1.362}_{-3.214}}$ & $-2.887^{+2.835}_{-4.762}$ & $-9.627^{+34.812}_{-67.386}$  & $1.437^{+0.107}_{-0.105}$ & $3.284^{+0.115}_{-0.112}$ & $0.604$ \\
\hline
$(0.25-0.5-1.0)$      & $0.273^{+0.021}_{-0.021}$ & $-1.352^{+3.806}_{-9.208}$ & $\mathbf{-1.253^{+0.271}_{-0.288}}$ & $\mathbf{-1.854^{+0.643}_{-1.883}}$ & $-2.610^{+0.861}_{-1.774}$ & $\mathbf{-3.356^{+2.721}_{-6.492}}$ & $-12.867^{+21.181}_{-43.511}$  & $1.438^{+0.114}_{-0.106}$ & $3.287^{+0.116}_{-0.110}$ & $0.540$ \\
\hline
$(0.25-0.75-1.0)$      & $0.268^{+0.020}_{-0.021}$ & $-1.016^{+8.047}_{-14.314}$ & $\mathbf{-1.233^{+0.252}_{-0.267}}$ & $-1.516^{+1.559}_{-2.948}$ & $\mathbf{-1.785^{+0.996}_{-2.021}}$ & $\mathbf{-2.024^{+1.682}_{-3.185}}$ & $-4.584^{+36.906}_{-62.470}$  & $1.433^{+0.108}_{-0.100}$ & $3.279^{+0.115}_{-0.105}$ & $0.743$ \\
\hline
$*(0.5-0.75-1.0)$       & $0.268^{+0.021}_{-0.020}$ & $-1.209^{+36.412}_{-59.684}$ & $-1.267^{+8.638}_{-14.354}$ & $\mathbf{-1.612^{+0.492}_{-0.926}}$ & $\mathbf{-2.009^{+1.273}_{-2.197}}$ & $\mathbf{-2.389^{+2.146}_{-3.532}}$ & $-7.023^{+78.529}_{-124.434}$  & $1.434^{+0.111}_{-0.104}$ & $3.287^{+0.112}_{-0.112}$ & $0.660$ \\
\hline
\hline
all $(2NL)$ & $0.271^{+0.021}_{-0.021}$ & $-1.115^{+3.231}_{-3.091}$ & $-1.241^{+1.105}_{-1.087}$ & $-1.782^{+1.316}_{-1.397}$ & $-2.417^{+2.146}_{-2.491}$ & $-3.026^{+3.829}_{-4.397}$ & $-10.145^{+52.542}_{-54.819}$  & $-$ & $-$ & $0.609$ \\
\hline
\hline
$(0-0.25-0.5-0.75)$      & $0.272^{+0.020}_{-0.021}$ & $\mathbf{-1.708^{+0.931}_{-1.024}}$ & $\mathbf{-1.374^{+0.313}_{-0.396}}$ & $\mathbf{-2.289^{+0.774}_{-1.204}}$ & $\mathbf{-1.540^{+1.289}_{-1.955}}$ & $0.539^{+3.984}_{-6.074}$ & $91.519^{+113.573}_{-6173.582}$  & $1.433^{+0.113}_{-0.103}$ & $3.302^{+0.113}_{-0.113}$ & $0.652$ \\
\hline
$(0-0.25-0.5-1.0)$       & $0.274^{+0.021}_{-0.022}$ & $\mathbf{-1.710^{+0.986}_{-1.125}}$ & $\mathbf{-1.376^{+0.333}_{-0.425}}$ & $\mathbf{-2.415^{+0.878}_{-1.544}}$ & $-2.082^{+1.893}_{-2.776}$ & $\mathbf{-0.596^{+4.548}_{-6.313}}$ & $74.042^{+109.162}_{-159.637}$  & $1.439^{+0.116}_{-0.107}$ & $3.310^{+0.118}_{-0.114}$ & $0.541$ \\
\hline
$(0-0.25-0.75-1.0)$       & $0.275^{+0.020}_{-0.021}$ & $\mathbf{-1.714^{+0.943}_{-1.011}}$ & $\mathbf{-1.368^{+0.326}_{-0.397}}$ & $-2.394^{+2.422}_{-4.025}$ & $\mathbf{-2.020^{+1.659}_{-2.990}}$ & $\mathbf{-0.475^{+4.209}_{-6.282}}$ & $75.851^{+203.264}_{-318.049}$  & $1.438^{+0.113}_{-0.108}$ & $3.302^{+0.113}_{-0.112}$ & $0.560$ \\
\hline
$*(0-0.5-0.75-1.0)$       & $0.275^{+0.021}_{-0.022}$ & $\mathbf{-1.705^{+0.984}_{-1.031}}$ & $-2.173^{+7.105}_{-20.538}$ & $\mathbf{-2.578^{+0.987}_{-1.847}}$ & $\mathbf{-2.029^{+1.710}_{-5.929}}$ & $\mathbf{-0.821^{+4.571}_{-11.312}}$ & $44.773^{+355.884}_{-996.800}$  & $1.437^{+0.113}_{-0.105}$ & $3.311^{+0.114}_{-0.116}$ & $0.555$\\
\hline
$*(0.25-0.5-0.75-1.0)$      & $0.266^{+0.019}_{-0.018}$ & $-1.719^{+21.325}_{-26.293}$ & $\mathbf{-1.161^{+0.206}_{-0.218}}$ & $\mathbf{-1.567^{+0.387}_{-0.477}}$ & $\mathbf{-1.731^{+0.821}_{-0.976}}$ &
 $\mathbf{-1.591^{+1.125}_{-1.568}}$ & $16.053^{+175.517}_{-220.443}$  & $1.438^{+0.113}_{-0.105}$ & $3.284^{+0.118}_{-0.110}$ & $0.860$ \\
\hline
\hline
all $(3NL)$ & $0.273^{+0.020}_{-0.021}$ & $-1.955^{+1.197}_{-5.961}$ & $-1.305^{+0.438}_{-0.544}$ & $-2.088^{+1.049}_{-1.545}$ & $-1.785^{+1.869}_{-2.041}$ & $0.628^{+3.950}_{-5.746}$ & $111.690^{+267.383}_{-225.100}$  & $-$ &
$-$ & $0.584$\\
\hline
\hline
\end{tabular}}
\end{minipage}
\end{table*}}}

\end{document}